\documentclass[twocolumn,showpacs]{revtex4}
\usepackage{epstopdf}
\usepackage{mathbbol}              
\usepackage{graphics,graphicx,epsfig,ulem}

\begin{document}

\title{Electro--optic control of atom--light interactions using Rydberg dark--state polaritons}
\date{\today}
\author{M. G. Bason, A. K. Mohapatra, K. J. Weatherill, and C. S. Adams
}
\affiliation{Department of Physics, Durham University,
Rochester Building, South Road, Durham DH1 3LE, England.}

\begin{abstract}              
We demonstrate a multiphoton Rydberg dark resonance where a $\Lambda$--system is coupled to a Rydberg state. This ${\cal N}$--type level scheme combines the ability to slow and store light pulses associated with long lived ground state superpositions, with the strongly interacting character of Rydberg states. For the $n{\rm d}_{5/2}$ Rydberg state in $^{87}$Rb (with $n=26$ or 44) and a beam size of 1~mm we observe a resonance linewidth of less than 100~kHz in a room temperature atomic ensemble limited by transit--time broadening. The resonance is switchable with an electric field of order $1$~Vcm$^{-1}$.
We show that, even when photons with different wavevectors are involved, the resonance can be Doppler-free. Applications in electro-optic switching and photonic phase gates are discussed.

\end{abstract}



\maketitle


As photons are the most robust carriers of quantum information there is considerable interest in developing single--photon sources, memories and gates. The storage and retrieval of single photons using
electromagnetically induced transparency (EIT) \cite{harris} in both laser cooled \cite{chan05}
and thermal atomic ensembles \cite{eisa05} have been demonstrated experimentally. These experiments employ the dark--state polariton concept where the photon is evolved into a single atomic spin excitation and back into a photon \cite{dsp}. Dark--state polaritons employ three atomic energy levels where two ground states are coupled via an excited state forming a $\Lambda$--system. The attractive feature of this level scheme is that the ground state coherence is long lived allowing photon storage times of order microseconds \cite{chan05,eisa05}.

In principal, deterministic photon gates could be realized using a large non--linearity at the single--photon level.
Enhanced Kerr non-linearities have been predicted in a medium displaying EIT \cite{schm96}, and have been observed in experiments on cold atoms \cite{kang03,wang06}. However, achieving a sufficiently large non--linearity and matched group velocities for both photon pulses \cite{luki00}, without introducing pulse distortion remains a problem. Recent theoretical work has focussed on possible solutions based on frequency and intensity tuning \cite{otta03}, tight confinement of the light pulse \cite{andr05}, enhancing the interactions using Rydberg states \cite{frie05} and double EIT \cite{wang06b}.
The strongly interacting character of Rydberg atoms \cite{gallagher} makes them particularly attractive for achieving entanglement, as discussed in the context of fast quantum gates for atoms \cite{jaksch00,lukin01}. Recently, we showed that it is possible to map the properties of Rydberg atoms, such as their high sensitivity to electric fields, onto a strong optical transition from the ground state using a ladder EIT scheme \cite{moha07}. Theoretical calculations suggest that this level configuration could produce a cross phase modulation of $\pi$ for counterpropagating single--photon pulses in a dense ultra--cold atomic ensemble providing a possible scheme for the realization of a photon gate \cite{frie05}.

In this paper, we propose and demonstrate a multiphoton Rydberg dark resonance where a $\Lambda$--system is coupled to a Rydberg state. This ${\cal N}$--type level scheme combines the attractive features of both ground state dark--state polaritons \cite{dsp} and Rydberg states, i.e., the ability to slow and store light pulses \cite{chan05,eisa05} and the strong interactions associated with Rydberg atoms \cite{gallagher}. For the $n$d$_{5/2}$ Rydberg state (with $n=26$ and 44) in $^{87}$Rb and a beam size of 1~mm we observe a resonance linewidth of less than 100~kHz in a room temperature vapor.
We show that the Rydberg dark resonance is switchable with an electric field of order $1$~Vcm$^{-1}$, and hence the strongly interacting properties of Rydberg states can be mapped onto the ground state coherence. Finally, we show that the Rydberg dark resonance can be Doppler--free creating the potential for the realization of efficient electro--optic single--photon devices using thermal atomic vapors.

The energy levels of $^{87}$Rb used for the experimental demonstration of the multi--photon Rydberg dark resonance are shown in Fig. 1(a). A $\Lambda$--system ($\vert 1\rangle\rightarrow \vert 2\rangle \rightarrow \vert 3\rangle$) is formed by a weak probe beam with Rabi frequency $\Omega_{\rm 21}$ resonant with
$5{\rm s}~^2{\rm S}_{1/2}(F=1) \rightarrow 5{\rm p}~^2{\rm P}_{3/2}(F'=2)$ transition
and a strong coupling beam with Rabi frequency $\Omega_{\rm 32}$ resonant with the $5{\rm s}~^2{\rm S}_{1/2}(F=2) \rightarrow 5{\rm p}~^2{\rm P}_{3/2}(F'=2)$ transition. The coupling beam is derived from a commercial $780.24$~nm extended cavity diode laser that is locked directly to the $F=2\rightarrow F'=2$ resonance using polarization spectroscopy \cite{pear02}. A fraction of the coupling beam is passed through an electro--optic modulator (EOM) that adds sidebands at 6.8~GHz. A slave laser is injection locked onto the upper sideband to obtain the probe beam. A high spectral purity is obtained by using a double injection scheme \cite{baso08}. The probe beam can be scanned through the $F=1\rightarrow F'=2$ resonance by changing the drive frequency of the EOM. The probe and coupling beams are combined on a polarization beamsplitting cube and coupled into a polarization maintaining single mode fiber. The output of the fibre is collimated and propagates through a room temperature Rb cell of length 75 mm as shown in Fig. 1(b). The probe and coupling beams have a waist of 1.3~mm ($1/{\rm e}^2$ radius) and powers of 10~$\mu$W and 35~$\mu$W, respectively.

The $\Lambda$--system is coupled to a Rydberg state using a 2--photon excitation scheme as shown in Fig. 1(a). The first step is formed by a 780.24 nm beam resonant with the $5{\rm s}~^2{\rm S}_{1/2}(F=2) \rightarrow 5{\rm p}~^2{\rm P}_{3/2}(F'=3)$ transition that is derived by shifting the coupling light using two acousto--optic modulators (AOMs). The laser power incident on the cell is 40~$\mu$W. The second step is derived from a commercial frequency doubled diode laser system (Toptica TA-SHG) at 480 nm that is resonant with $5{\rm p}~^2{\rm P}_{3/2}(F'=3) \rightarrow n{\rm d}~^2{\rm D}_{5/2}(F'')$ transition (the $n$d state hyperfine splitting is not resolved).
The 2--photon resonance is driven by counter--propagating beams in the vapor cell with the 2nd step co--propagating with the $\Lambda$-system beams as in Fig. 1(b). The 480~nm beam has a power of 140~mW and a waist of 0.8~mm ($1/{\rm e}^2$ radius). The waist size is a trade--off between mode--matching with the 780~nm beams and achieving a high Rabi frequency. The Rb vapor cell
is placed between two copper bar electrodes and inside a single layer magnetic shield.
The 480~nm laser is frequency stabilized using a ladder system Rydberg EIT signal \cite{moha07} in a second Rb vapor cell. From the EIT error signal the line width of the 2--photon resonance is $\sim 150$~kHz.

\begin{figure}[]
\begin{center}
\epsfig{file=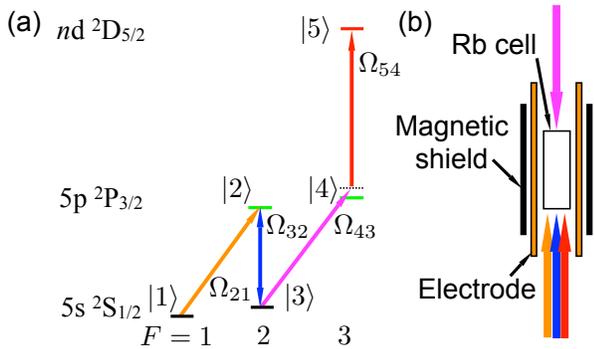,clip=,width=8.0cm}
\caption[]{(a) Level scheme used for the experimental demonstration of a 4--photon Rydberg dark resonance:
The $\Lambda$--system ($\vert 1\rangle \rightarrow \vert 2\rangle\rightarrow \vert 3\rangle$) is coupled to a Rydberg state, $\vert 5\rangle$ using a 2--photon transition via an intermediate state, $\vert 4\rangle$. The absorption or dispersion
of the medium is detected on the transition $\vert 1\rangle \rightarrow \vert 2\rangle$ using a probe beam with Rabi frequency
$\Omega_{\rm 21}$.  (b) Schematic of the experimental set--up. The $\Lambda$--system beams (orange and blue) co--propagate with the second step of the Rydberg excitation (red). The first step of the Rydberg excitation (purple) is counter--propagating.}
\end{center}
\end{figure}

The spectrum corresponding to the $\Lambda$--system resonance is shown in Fig.~2(a). The line width of $\sim 70$~kHz is limited mainly by transit time broadening. To isolate the effect of the Rydberg coupling on the $\Lambda$--system we modulate the amplitude of the 480 nm laser at 35~kHz and monitor the probe signal using a lock--in amplifier.
The lock--in signal which corresponds to the change in the $\Lambda$ resonance when the Rydberg coupling is switched on and off is shown in Fig.~2(b). This signal disappears when any one of the four beams involved is blocked.

\begin{figure}[]
\begin{center}
\epsfig{file=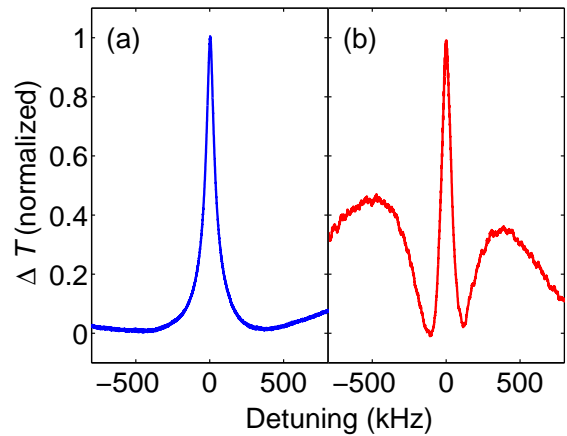,clip=,width=7.5cm}
\caption[]{Experimental measurement of the change in transmission, $\Delta T$, associated with (a) the $\Lambda$--system, $\vert 1\rangle \rightarrow \vert 2\rangle \rightarrow \vert 3\rangle$ and (b) the Rydberg dark resonance signal corresponding to the difference between the $\Lambda$--system resonance with Rydberg coupling ($n=26$) on and off. The Rydberg dark resonance is measured with a lock--in amplifier and is two order of magnitude smaller than the $\Lambda$--system resonance.}
\end{center}
\end{figure}

\begin{figure}[]
\begin{center}
\epsfig{file=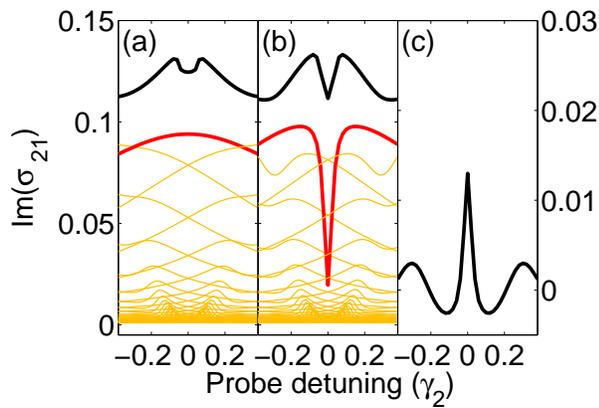,clip=,width=8.0cm}
\caption[]{The absorption (proportional to the imaginary part of the density matrix component, $\sigma_{21}$) experienced by the probe beam as a function of detuning for a 4--photon Rydberg dark resonance with the Rydberg coupling (a) off and (b) on. In (c) we show the difference between (a) and (b), which corresponds to the effect of coupling to the Rydberg state as observed in the experiment. The specific parameters are:
$\Omega_{\rm 21}=0.02\gamma_2$, $\Omega_{\rm 32}=0.2\gamma_2$, $\Omega_{\rm 43}=0.8\gamma_2$, $\Delta_{43}=\Delta_{54}=0$, $\gamma_4=\gamma_2$ and $\gamma_5=0.01\gamma_2$. (a) $\Omega_{\rm 54}=0.0\gamma_2$ and (b) $\Omega_{\rm 54}=0.8\gamma_2$. The thin lines show velocities in the range $-10\gamma_2\lambda_1$ to $+10\gamma_2\lambda_1$ in steps of $0.5\gamma_2\lambda_1$. The thick red (grey) lines indicates the absorption of the zero velocity class atoms. The thick black line indicates the total absorption obtained by summing over velocity. The curves for each velocity class are multiplied by 5 so that they can be plotted on the same scale as the sum.
}
\label{fig:theory1}
\end{center}
\end{figure}

To understand the shape of the observed Rydberg dark resonance signal and establish how the effect may be optimized in both thermal and ultra--cold atomic ensembles we have solved the optical Bloch equations for the 5--level system. The 5--level Hamiltonian may be written as
$$
H=\frac{1}{2}\left(\begin{array}{ccccc}
\Delta_{21} & \Omega_{21} & 0 & 0 & 0\\
\Omega_{21} & \Delta_{32}-k_1v & \Omega_{32} & 0 & 0\\
0 & \Omega_{32} & 0 & \Omega_{43} & 0\\
0 & 0 & \Omega_{43} & \Delta_{43}+k_1 v & \Omega_{54}\\
0 & 0 & 0 & \Omega_{54} & \Delta_{54}+k'v
\end{array}\right)
$$
where the labels 1 to 5 refer to the states $5{\rm s}~^2{\rm S}_{1/2}(F=1)$, $5{\rm p}~^2{\rm P}_{3/2}(F'=2)$, $5{\rm s}~^2{\rm S}_{1/2}(F=2)$, $5{\rm p}~^2{\rm P}_{3/2}(F'=3)$ and  $44{\rm d}~^2{\rm D}_{5/2}$, respectively, as in Fig. 1(a), $\Delta_{ij}$ and $\Omega_{ij}$ are the detuning and Rabi frequency between states $i$ and $j$, $k_1$ and $k_2$ are the wavevectors of the 780 and 480 nm beams, $k'=k_1-k_2$, and $v$ is the atomic velocity. The population in state $i$ and coherence involving state $i$ decay at a rate $\gamma_i$ and $\gamma_i/2$, respectively. Note that this is not a complete description of the experiment as the hyperfine splitting of the $5{\rm p}~^2{\rm P}_{3/2}$ state is less than the Doppler broadening so there are multiple contributions to both the $\Lambda$ and two--photon Rydberg resonances. However, the model does give qualitative insight into the properties of the Rydberg dark resonance. In Fig.~3 we plot the imaginary part of the coherence between states $\vert 1\rangle$ and $\vert 2\rangle$ which determines the probe absorption. The thin lines indicate the absorption for different velocity classes. The thick red lines indicates the absorption of the zero velocity class atoms. The thick black line indicates the total absorption obtained by summing over velocity. Fig.~3(c) shows the difference between the cases where the 480~nm laser is on and off, and reproduces the shape the observed Rydberg dark resonance seen in Fig.~2(b). Fig.~\ref{fig:theory1}(a) and (b) show that the narrow contribution to the Rydberg dark resonance is from the zero velocity class atoms, while the wings of the signal arise from low velocity atoms.

The interesting feature of the ${\cal N}$--system Rydberg dark resonance is the ability to map the strongly interacting character of Rydberg atoms onto a ground state coherence. To demonstrate this we show that as a result of the Rydberg coupling, the $\Lambda$--system resonance has become highly sensitive to electric fields.
We apply an ac electric field with frequency 20~MHz to the copper bar electrodes shown in Fig. 1(b), and measure the amplitude of the central peak of the Rydberg dark resonance signal as a function of the rms field. An ac field is used to avoid the effects of charge screening inside the cell \cite{moha07}. The results are shown in Fig.~4. The Rydberg dark resonance is suppressed with a field of order $1$~Vcm$^{-1}$. The shape of the curve can be explained by the electric field dependence of the $44{\rm d}~^2{\rm D}_{5/2}$ state which splits into three components with the $\vert m_J\vert =5/2$ component shifting out of resonance at a lower field (of order 0.1~Vcm$^{-1}$) and the other two components shifting out of resonance at 0.9~Vcm$^{-1}$ \cite{moha07}. As the polarization scales as the seventh power of the principal quantum number, $n^7$, an even higher sensitivity can be expected for Rydberg states with larger $n$.

\begin{figure}[]
\begin{center}
\epsfig{file=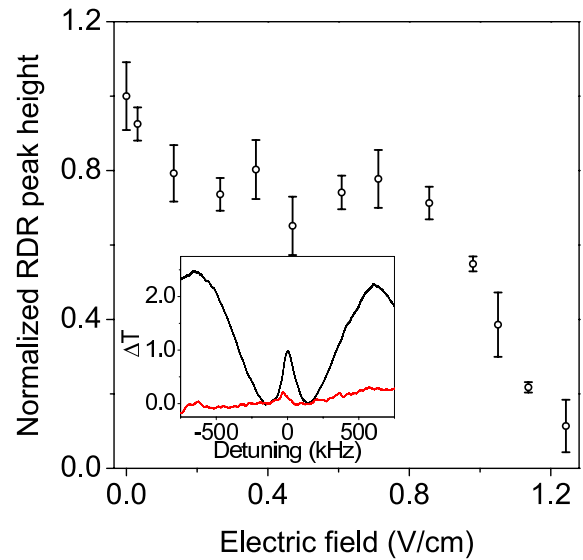,clip=,width=7.5cm}
\caption[]{Electric field dependence of the Rydberg dark resonance (RDR) amplitude for $n=44$. The RDR signals at  0 and 1.1 Vcm$^{-1}$ are shown inset. Note that the relative height of the central peak and the wings is smaller for $n=44$ than $n=26$, Fig. 2(b), due to the lower value of $\Omega_{54}$.}
\end{center}
\label{fig:rdr_efield}
\end{figure}

As the narrow central peak of the Rydberg dark resonance arises from the zero velocity class atoms, the signal could be enhanced either by using cold atoms or if the multi--photon resonance could be made Doppler--free. A Doppler--free configuration in a thermal ensemble would be particularly attractive due to the reduced experimental overhead involved in state preparation, resulting in much higher `clock' rate for a practical device.
By examining the eigenvalues of the 5--level Hamiltonian one finds that if $\Delta_{43}\gg \gamma_2$, such that the 2--photon transition to the Rydberg state becomes an effective single--photon transition with Rabi frequency $\Omega_{53}=\Omega_{43}\Omega_{54}/2\Delta_{43}$ with wavevector $k'$ co--propagating with the $\Lambda$ beams, and if one balances the Rabi frequencies of the two coupling transitions $\Omega_{\rm 53}/\Omega_{\rm 32}=(k'/k_1)^{1/2}$, one finds a Doppler--free solution, i.e., an eigenvalue that is independent of $v$. Previously, Doppler--free dark states have only been identified in a 4--level system for the case where all the $k$ vectors are equal \cite{ye02}.
The absorption properties of the medium for such a Doppler--free Rydberg dark are illustrated in Fig.~5. Since the resonance is Doppler--free, all velocity classes contribute to the total absorption or transparency and the effect of the Rydberg coupling is dramatic. We choose to show the effect of a shift in the Rydberg energy level that could arise due to an external electric field or Rydberg--Rydberg interactions. When the Rydberg coupling is on--resonance, Fig. 5(a), the medium is strongly absorbing at $\Delta_{21}=-0.2\gamma_2$. But if the Rydberg state is shifted out of resonance, Fig. 5(b), the medium becomes completely transparent at this detuning.
This means one could use an applied electric field to switch the media from opaque to transparent forming an electro--optic shutter.
Similarly, Rydberg--Rydberg interactions that give rise to the dipole blockade mechanism \cite{sing04,tong04,cube05,pillet06,heid07} would modify the medium from absorbing with a superluminal group velocity for single excitation, to transparent with a slow group velocity for multiple excitations, resulting in a large non-linearity at the single--photon level.

There are a number of additional points of note in this example: First
the sensitivity to the Rydberg energy level, i.e., $\Delta_{54}$, depends on the decoherence rate of both levels 2 and 5, so ones requires $\Delta_{54}>\gamma_2$ to change the properties of the medium. However, this is easily satisfied for Rydberg states as the polarizability and Rydberg--Rydberg interactions can be large. For example, for an  $n=80~$s--state in Rb, the van der Waals shift between two Rydberg atoms is of order $50\gamma_2$ at an atomic spacing of $5~\mu$m \cite{sing05}.
Second, the detuning from the intermediate state in the two--photon transition must be larger than the Doppler width which means that relatively high single photon Rabi frequencies are required, i.e., $\Omega_{43}$ and $\Omega_{54}\gg\gamma_2$. This could be achieved in a thermal vapor by focusing the beams in a short cell.
Finally, the first step of the two--photon excitation of the Rydberg state induces an ac--Stark shift such that the resonance is slightly asymmetric and shifted to non--zero probe detuning. This can be avoided by making $\Omega_{54}>\Omega_{43}$.

\begin{figure}[]
\begin{center}
\epsfig{file=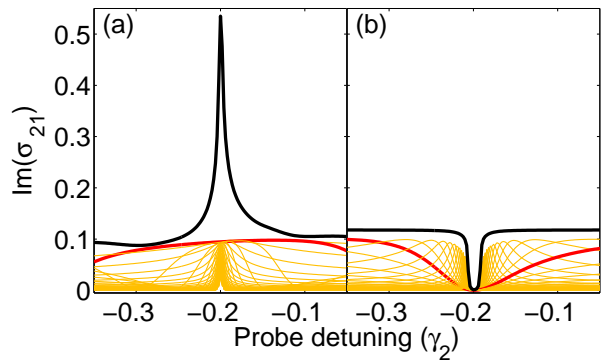,clip=,width=8.0cm}
\caption[]{Example of a Doppler--free 4--photon Rydberg dark resonance where in (a) the Rydberg coupling is resonant ($\Delta_{54}=0$), while in (b) the Rydberg energy level is shifted by an external electric field or atomic interactions such that $\Delta_{54}=50\gamma_2$. The other parameters are:
$\Omega_{\rm 21}=0.02\gamma_2$, $\Omega_{\rm 32}=0.25\gamma_2$, $\Omega_{\rm 43}=\Omega_{\rm 54}=20\gamma_2$, $\Delta_{43}=2000\gamma_2$, $\gamma_4=\gamma_2$ and $\gamma_5=0.01\gamma_2$. The thin lines show velocities in the range $-10\gamma_2\lambda_1$ to $+10\gamma_2\lambda_1$ in steps of $0.5\gamma_2\lambda_1$.}
\end{center}
\end{figure}

In summary, we have demonstrated a 4--photon Rydberg dark resonance with a linewidth of less than 100~kHz limited mainly by transit--time broadening. We show that resonance can be suppressed by applying an electric field of order $1$~Vcm$^{-1}$, and thereby demonstrate that the strongly interacting properties of Rydberg states can be mapped onto the ground state coherence. We show that the resonance can be made Doppler--free, allowing electro--optic switching in a thermal vapor.



We are grateful to MPA Jones for stimulating discussions. We thank the EPSRC for financial support.

\end{document}